\pdfoutput=1
\documentclass[11pt]{article}
\usepackage[]{acl}

\usepackage{times}
\usepackage{latexsym}
\usepackage[pdftex]{graphics}

\usepackage[T1]{fontenc}
\usepackage[utf8]{inputenc}

\usepackage{microtype}

\usepackage{algorithm}
\usepackage{algpseudocode}
\usepackage{multirow}
\usepackage{amsmath}
\usepackage{booktabs}
\usepackage{epstopdf}
\usepackage{float}
\usepackage{times}
\usepackage{latexsym}
\usepackage{graphicx}

\title{HLATR: Enhance Multi-stage Text Retrieval with \\ Hybrid List Aware Transformer Reranking}

\author{Yanzhao Zhang, Dingkun Long, Guangwei Xu, Pengjun Xie \\
  Alibaba Group \\
  \texttt{zhangyanzhao.zyz,dingkun.ldk,kunka.xgw,pengjun.xpj@alibaba-inc.com} \\
}  

\begin{document}
\maketitle
\begin{abstract}
Deep pre-trained language models (e,g. BERT) are effective at large-scale text retrieval task. Existing text retrieval systems with state-of-the-art performance usually adopt a retrieve-then-reranking architecture due to the high computational cost of pre-trained language models and the large corpus size. Under such a multi-stage architecture, previous studies mainly focused on optimizing single stage of the framework thus improving the overall retrieval performance. However, how to directly couple multi-stage features for optimization has not been well studied. In this paper, we design Hybrid List Aware Transformer Reranking (HLATR) as a subsequent reranking module to incorporate both retrieval and reranking stage features. HLATR is lightweight and can be easily parallelized with existing text retrieval systems so that the reranking process can be performed in a single yet efficient processing. Empirical experiments on two large-scale text retrieval datasets show that HLATR can efficiently improve the ranking performance of existing multi-stage text retrieval methods\footnote{Our code will be available at https://github.com/Alibaba-NLP/HLATR}.
\end{abstract}

\section{Introduction}
Text retrieval is a task which aims to search in a large corpus for texts that are most relevant given a query. The final retrieval result is informative and can benefit a wide range of down streaming applications, such as open domain question answering~\cite{karpukhin2020dense,Li2021EncoderAO}, machine reading comprehension~\cite{Rajpurkar2016SQuAD1Q,Nishida2018RetrieveandReadML}, and web search systems~\cite{huang2020embedding,liu2021pre}.

\begin{figure}[t]
    \centering
    \includegraphics[width=1.0\columnwidth]{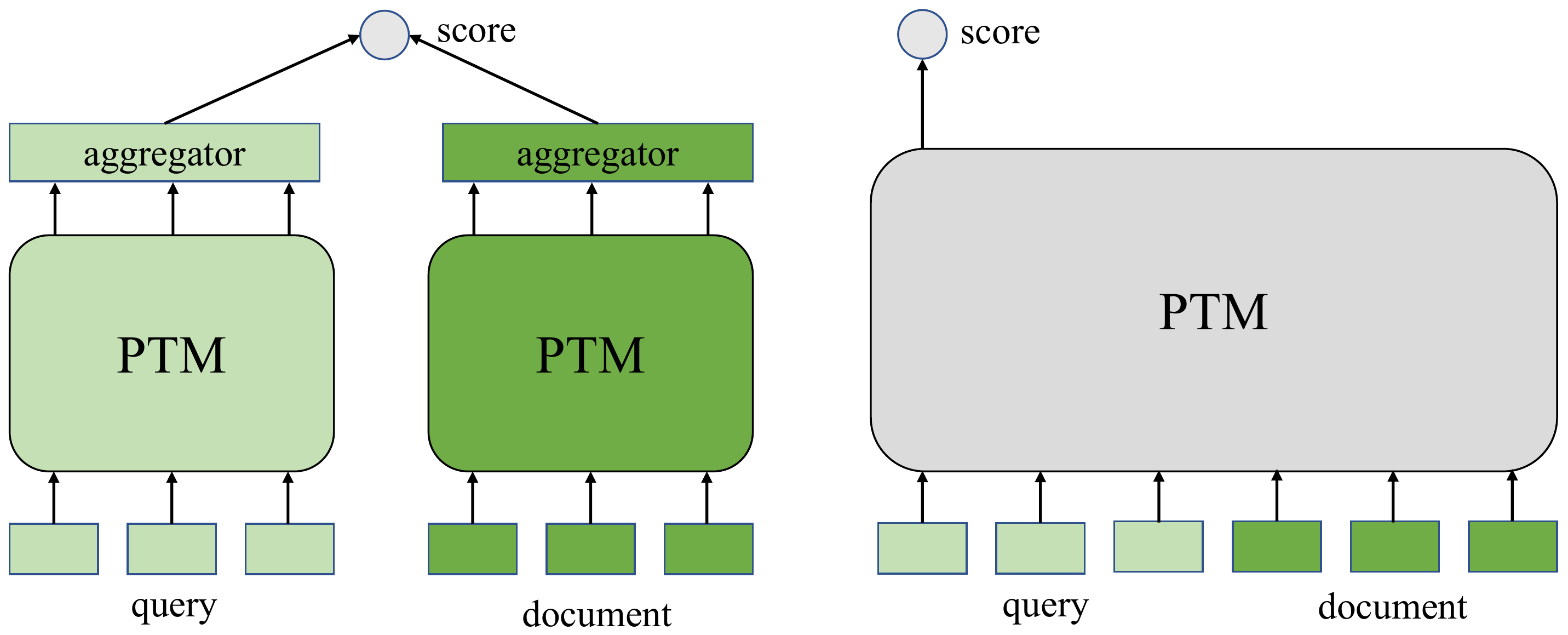} 
    \caption{Model architectures of representation-focused model (left) and interaction based model (right).}
    \label{fig:two-type-model}
\end{figure}

Since the corpus is massive in general, a practical text retrieval system usually leverages a retrieve-then-reranking architecture due to the efficiency and effectiveness trade-off. Specifically, given a search query, a retriever is involved to efficiently recall a relatively small size of relevant texts from the large corpus. Then the query and the initial retrieval results will be passed through more delicate reranking stages to produce the final ranked results. In recent years, the emergence of deep pre-trained language models (PTM)~\cite{kenton2019bert,liu2019roberta} has brought a storm in natural language processing (NLP) and information retrieval fields. Text retrieval systems armed with pre-trained language models have become a dominant paradigm to improve the overall performance compared to traditional statistical methods (e,g. BM25). Considering the large computational cost of transformer-based PTMs, the retrieval stage and reranking stage usually use different model architectures: representation-focused model and interaction-focused model respectively~\cite{fan2021pre}, as illustrated in Figure \ref{fig:two-type-model}.

Based on the model structures presented in Figure \ref{fig:two-type-model}, previous approaches to multi-stage text retrieval roughly fall into two groups. The first group tend to separately optimize the retrieval or reranking stage by designing task-oriented PTMs~\cite{gao2021condenser,gao2021unsupervised,ma2021b} or better model training strategies~\cite{xiong2021approximate,gao2021rethink}. Studies in the second group attend to jointly optimize the two stages by adversarial training~\cite{zhang2021adversarial} or knowledge distillation~\cite{ren2021rocketqav2,lu2022erniesearch}. However, this kind of joint learning method mainly focused on improving the retrieval stage performance by absorbing merits from the reranking stage.

In a global view, the retrieval stage and reranking stage are highly correlated. Intuitively, better retrieval results will definitely provide a more enlightening signal to the reranking stage~\cite{gao2021rethink} for both the training and inference phase. Furthermore, since the two individual stages adopt different model architectures (representation-focused vs interaction-focused), we infer that the focus of the two stages is also different despite the sharing target to estimate the relevance of query and document. Empirically, we find that simply a weighted combination of the relevance scores produced by the two stages can further improve the overall performance, which also supports our hypothesis above. These observations inspire us whether it is possible to make an additional module that couples both coarse retrieval and fine reranking features to facilitate the final retrieval performance.

In this paper, we propose to add an additional reranking stage with a lightweight reranking model, which we noted as Hybrid List Aware Transformer Ranker (HLATR). As we tend to incorporate both the two-stage features and the reranking training objective is list aware, we adopt the transformer model structure~\cite{vaswani2017attention} for HLATR. As introduced in~\cite{vaswani2017attention}, the content for each position of the sequential inputs includes the embedding itself and the position embedding corresponded. Here, we carry out the feature coupling process by taking the logit produced by the PTM based reranking model as the embedding of each document, and we compute the position embedding based on the retrieval ranking order of each document in the retrieval stage. A stack of transformer layers will be used as the encoder part. In HLATR, we can effectively fuse the retrieval and reranking features and model the list aware text ranking process. 

To verify the effectiveness and robustness of HLATR model, we conduct experiments on two commonly used large text retrieval benchmarks, namely the MS MARCO passage and document ranking datasets~\cite{Campos2016MSMA}. Empirically experiment results demonstrate that HLATR can remarkable improve the final text retrieval performance with different combinations of retrieval and reranking architectures. Further we design sufficient subsidiary experiments to prove that HLATR is lightweight and easy to practical.

\section{Related Work}
In order to balance efficiency and effectiveness, existing text retrieval systems usually leverage a retrieval-then-reranking framework. 

For the retrieval stage, traditional sparse retrieval methods adopt the exact term matching signal for text relevance estimation (e,g. TF-IDF~\cite{luhn1957statistical} and BM25~\cite{robertson2009probabilistic} method). Recently, with the rapid development of the deep text representation method, many works shifted to training dense retrievers by taking BERT or other deep neural networks as text encoders~\cite{karpukhin2020dense,nogueira2019passage}. These models utilize dual encoders to encode queries and documents separately, which allows pre-encoding and indexing of corpus. Under this paradigm, various optimization strategies have been proposed to improve the retrieval performance including designing task-specific pre-trained model~\cite{gao2021unsupervised,gao2021condenser}, hard negatives mining strategy~\cite{xiong2021approximate} and multi-view text representation~\cite{hu2021multi}.

For the reranking stage, previous studies use an interaction-focused model which takes the pair of query and document as input. By considering the token level similarity, these models always have a better ranking power~\cite{guo2016deep}. By such a full interaction computation strategy, the PTM based reranking model usually reranks a small number of documents produced by the retrieval stage. ~\cite{ma2021b} propose to enhance the pre-trained language model by constructing scale pseudo relevant query-document pairs.  ~\cite{gao2021rethink} propose to replace the traditional binary cross-entropy loss~\cite{nogueira2019passage} with a contrastive loss for better list aware ranking. ~\cite{pu2021yes} attend to dynamically select hard negative samples to construct high-quality semantic space the trained reranking model.

Apart from the separate optimization methods presented above, there are also studies seeking to jointly optimize the retrieval and reranking stage. In ~\cite{ren2021rocketqav2}, they simultaneously optimize the retrieval model and the ranking model with a hybrid data augmentation technology. AR2~\cite{zhang2021adversarial} uses a Generative adversarial network (GAN)~\cite{creswell2018generative} method which regards the retrieval model as the ``generative mode'' and the ranker model as the ``discriminative model''. All of these jointly learning methods need a large extra training cost than the traditional two-stage training method. In contrast, HLTAR requires only a very small additional consumption cost.

\begin{figure*}
    \centering
    \includegraphics[width= 2.0\columnwidth]{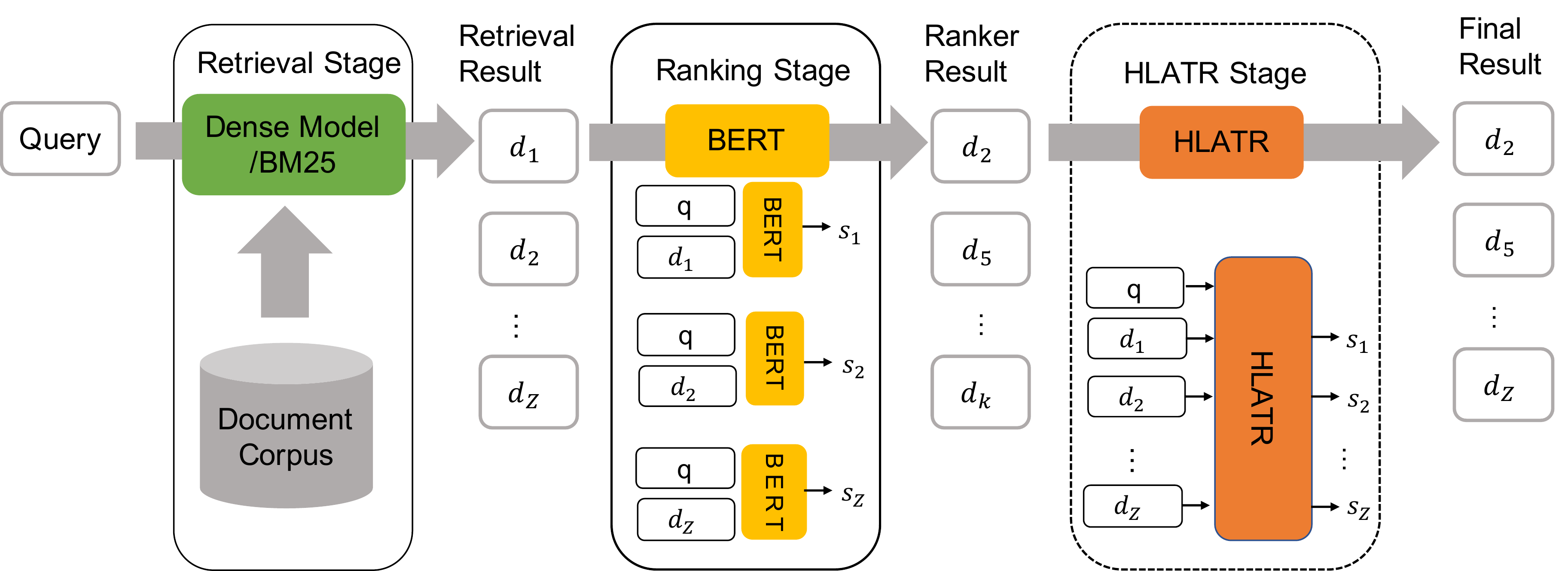}   
    \caption{An illusion of the conventional retrieve-then-reranking framework (solid line part) and our HLTAR ranking model (solid line and dashed line part).}
    \label{fig:frame}
\end{figure*}

\section{Methodology}

\subsection{Problem Overview}
The retrieve-then-reranking framework for text retrieval task is presented in Fig~\ref{fig:frame}. Given a query $q$, a text retrieval system aims to return a small size list of relevant documents from a large document set $\mathcal{D}=\{d_i\}_{i=1}^N$. Existing retrieval systems usually consist of a retrieval stage and a reranking stage with different focuses on efficiency and effectiveness. Here, for simplicity, we only introduce the dense retrieval model and reranking model based on pre-trained language model (as illustrated in Figure~\ref{fig:two-type-model}).

The retrieval stage takes $q$ and $\mathcal{D}$ as input and aims to return all potentially relevant documents $\mathcal{D}_r \in \mathcal{D}$ refer to the relevance between query and document, where $|\mathcal{D}_r| \ll N$. Dense retrieval model focuses on leaning to map text into quality continuous vector representation. By taking the pre-trained language model as the encoder, the retrieval model can be formulated as:
\begin{equation}
\label{eq:retrieval}
    score(q,d) = f(E^Q(q), E^D(d)),
\end{equation}
where $E^Q$ and $E^D$ are query and document encoders, and $f$ is the relevance estimation function.

Following the retrieval stage, a reranking stage takes $q$ and $\mathcal{D}_r$ as input to perform a more refined relevance evaluation for each query-document pair $(q, d)$, where $d \in \mathcal{D}_r$. Commonly, the reranking model adopts the interaction-based model. Without loss of generality, the reranking method could be abstracted by the following formulation:
\begin{equation}
\label{eq:reranking}
    score(q,d) = f(E^R(q,d)),
\end{equation}
where $E^R$ notes the PTM based encoder and $f$ is the evaluation function that computes the relevance score. 

In the retrieval stage, query and document are encoded independently. Therefore, the embedding representation of $\mathcal{D}$ can be pre-encoded. While in the reranking stage, the final representation of each query-document pair can not be pre-encoded due to the interaction-based architecture. As the result, we can only rerank a small set of documents as a result of the large computational cost of PTM. 

Despite adopting different model architectures, the retrieval and reranking stage share the same target for evaluating the relevance of query and document. Certainly, the training of retrieval and reranking models relies on large-scale labeled data. Previous studies~\cite{karpukhin2020dense,gao2021rethink} proposed to optimize the PTM based retrieval and reranking model with a contrastive learning objective. Concretely, for each query $q$, we form a group $G_d$ with a single relevant positive document $d^+$ and multiple negatives. By taking the scoring function defined in equations (\ref{eq:retrieval})(\ref{eq:reranking}), the contrastive loss for each query $q$ can be denoted as:
\begin{equation}
    \mathcal{L}_q := -log\frac{exp(score(q, d^{+}))}{\sum_{p\in G_d} exp(score(q, d))}.
\end{equation}

\subsection{HLATR}
\begin{figure}[t]
    \centering
    \includegraphics[width= 1.0\columnwidth]{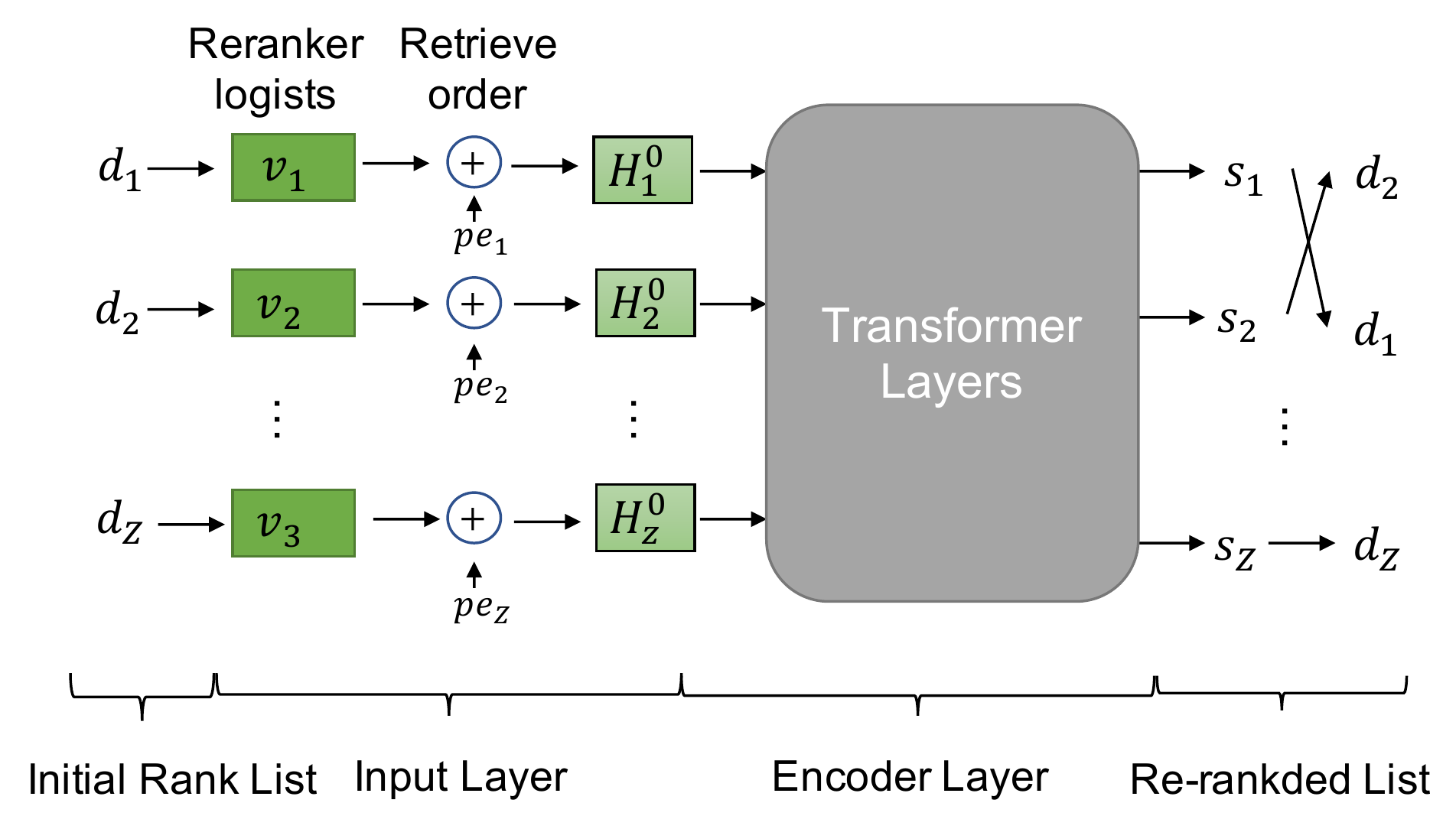}   
    \caption{Thee detailed network structure of our HLATR (Hybrid List Aware Transformer Reranking) model.}
    \label{fig:hltar}
\end{figure}

We target on building a lightweight reranking module that couples feature from both the retrieval and reranking stage. The overall architecture of our proposed model is illustrated in Figure \ref{fig:hltar}. Specifically, our model consists of a feature combination layer along with a transformer encoder. 

\subsubsection{Multi-stage Feature Fusion}
As the HLATR model is pipelined with the previous reranking model, HLATR takes the query $q$ and $D_r^{'}$ as input, where $D_r^{'}$ is the ranking results produced by the PTM based reranking module. Obviously, $|D_r| = |D_r^{'}| = Z$, where $Z$ denotes the reranking size. Since we adopt the transformer model architecture as the encoder, we carry out the multi-stage features fusion via coupling the retrieval ranking order in the retrieval stage and the final logit representation generated by the reranking model of each document.

For document $d\in D_r^{'}$ with retrieval order $i$ (where $0 \leq i \leq Z$), we map the $i$-th order into a learnable position embedding $\mathbf{pe_{i}} \in {\mathbf{R}^{d}}$ with dimension $d$, and which will be jointly optimized during the learning process. The position embedding is randomly initialized. The strategy of introducing retrieval order as a feature makes our model not only practicable for systems that use dense retrieval models, but also for others based on traditional statistical retrieval methods (e.g, BM25).

For the reranking stage, we hire the final encoded representation $E^R(q,d)$, which is converted to an embedding representation sharing the same dimension as the position embedding:
\begin{equation}
    v_i = W_{v}^TE^R(q,d_i),
\end{equation}
where $W_{v}$ is the trainable projection matrix.

Then, for each $d_i$ in $D_r^{'}$, we can directly add $pe_{i}$ and $v_i$ to build the input matrix $H^0 \in R^{d \times Z}$:
\begin{align*}
    H^0 = \begin{pmatrix} LN(pe_{1} + v_1) \\  LN(pe_{2} + v_2) \\  ... \\  LN(pe_{Z}+v_Z)
    \end{pmatrix},
\end{align*}
where LN represents the LayerNorm~\cite{ba2016layer} operation.

\subsubsection{Transformer Encoder} 
The learning process of the reranking model is list aware. The key point of a reranking module is to model and learn the partial order relationship between different documents within the entire candidates pool. Thus, we adopt the transformer architecture~\cite{vaswani2017attention} as the encoder owing to its long-distance modeling ability and full attention mechanism. The self-attention mechanism in the transformer can directly model the mutual influences for any different query-documents pair sample.

The entire encoder module consists of $L$ transformer layers. The hidden size of each Transformer layer is set equal to the input dimension $d$. For the first Transformer layer, it takes $H^0$ as input. While for the $l$-th Transformer layer $T^{l}$, it takes the output of the previous layer as input:
\begin{align*}
    H^{1} = T^1(H^0), \\
    H^{l} = T^l(H^{l-1}).
\end{align*}

\subsubsection{Learning Objective}
\label{sec:loss}
Finally, we use a linear projection layer to predict the relevance score between the input query $q$ and $D_r^{'}$. For each $d_i \in D_r^{'}$, the final relevance score can be represented as: 
\begin{equation}
    score(q, d_i) = W^TH^l_i.
\end{equation}
Similar to the previous stage,  we optimize HLATR with a list-wise contrastive loss~\cite{gao2021rethink}:
\begin{equation}
   \mathcal{L}_{\{q, D_r^{'}\}} = -log\frac{exp(score(q, d^{+}))}{\sum_{d\in D_r^{'}} exp(score(q, d))} .
\end{equation}

From the perspective of the learning objective, HLATR can also be classified as a reranking model, which takes a query $q$ and a document list as input and then produces the relevance score of each query-document pair among all candidate documents. HLATR is different from the previous reranking model based on the pre-trained language model. We sort out the difference between the two modules from three aspects:

\noindent 1) Superficially, the input of HLATR and the PTM based reranking model is the query $q$ and its corresponding candidate document list. Indeed, the PTM based reranking model uses the raw content of query and document as input, but HLATR uses the combination of features generated by the previous two stages as input. 

\noindent 2) Moreover, the calculation paradigm of the final ranking score is also different. In the PTM based reranking model, the relevance score of each query-doc pair is calculated independently. In the HLATR model, the relevance score of the entire input document list is calculated at one time through the serialization computing mechanism of the transformer architecture. Such a list aware calculation method allows us to better model the mutual relationship between the different documents in the entire input document list.

\noindent 3) During the training process, for PTM based reranking model, we have to control the data size of each mini-batch via a random sampling strategy when constructing $G_q$ for each query $q$ due to the high computational cost of PTM. The lightweight nature of HLATR allows us to use the entire retrieval results to construct the set of negatives, thereby greatly expanding the number of negative samples in each mini-batch. A previous study has proven that the larger number of negative samples can effectively improve the reranking performance under the contrastive learning optimization objective~\cite{gao2021rethink}.

\section{Experiment}
\subsection{Setup}

\noindent \textbf{Dataset} In this paper, we conduct experiments on two widely-used large-scale text retrieval datasets: MS MARCO Passage (called ``Passage'' below for simplicity) dataset and MS MARCO Document (called ``Document'' below) dataset. The MS MARCO dataset is constructed from Bing\'s search query logs and web documents retrieved by Bing. The statistic information of MS MARCO dataset is presented in Table~\ref{tab:static}. Since the number of parameters of HLATR is much smaller compared with the PTM reranking model, we only need to use a small amount of data to train the HLATR model. For the passage dataset, there exists an extra dev dataset which contains $34,000$ queries which can be used as training dataset. For the document dataset, we split $50,000$ queries from the orinal training dataset as the final training dataset for HLATR.

\vspace{0.15cm}
\noindent \textbf{Evaluation Metric} Following the evaluation methodology used in previous work, we report MRR@10 and MRR@100 results for the MS MARCO dev dataset due to the label of test data is not available. 

\begin{table}[]
\caption{Statistics of MA MARCO dataset. ``\#Length'' represents the average text length and ``\#Relvant'' represents the number of the labeled relevant documents per query. We only count the query which at least has one relevant document.}
\label{tab:static}
\begin{tabular}{@{}c|c|c@{}}
\toprule
Dataset             & Document & Passage \\ \midrule
\#Query             & 367013/5193               & 502939/6980  \\
\#Document          & 3213835                   & 8841823  \\
\#Length            & 56.8                        &  1127.8     \\
\#Relevant  & 1.04                        & 1.06       \\ \bottomrule
\end{tabular}
\end{table}

\begin{table*}[t]
\caption{MRR@10 and MRR@100 metrics for MS MARCO Passage and Document Dataset with different kinds of retrieval and reranking models. We bold the best performances of each combination of retrieval and reraking model. The results of HLATR are statistically significant difference ($p$ < $0.01$) compared to other models.}
\label{tab:res}
\resizebox{2.0\columnwidth}{!}{
\begin{tabular}{@{}cccccccccc@{}}
\toprule
\multicolumn{2}{c}{Model/Dataset} &
  \multicolumn{4}{c}{Passage} &
  \multicolumn{4}{c}{Document} \\ \midrule
\multicolumn{1}{c|}{\multirow{2}{*}{PTM}} &
  \multicolumn{1}{c|}{\multirow{2}{*}{}} &
  \multicolumn{2}{c|}{Sparse} &
  \multicolumn{2}{c|}{Dense} &
  \multicolumn{2}{c|}{Sparse} &
  \multicolumn{2}{c}{Dense} \\ \cmidrule(l){3-10} 
\multicolumn{1}{c|}{} &
  \multicolumn{1}{c|}{} &
  MRR@10 &
  \multicolumn{1}{c|}{MRR@100} &
  MRR@10 &
  \multicolumn{1}{c|}{MRR@100} &
  MRR@10 &
  \multicolumn{1}{c|}{MRR@100} &
  MRR@10 &
  MRR@100 \\ \midrule
\multicolumn{2}{c|}{Retrieval Only}  & 
  18.7 &
  \multicolumn{1}{c|}{19.8} &
  38.3 &
  \multicolumn{1}{c|}{39.4} &
  20.9 &
  \multicolumn{1}{c|}{22.2} &
  36.8 &
  37.9 \\ \midrule
\multicolumn{1}{c|}{\multirow{3}{*}{BERT$_{\text base}$}} &
  \multicolumn{1}{c|}{Reranking} &
  34.4 &
  \multicolumn{1}{c|}{35.5} &
  40.1 &
  \multicolumn{1}{c|}{41.1} &
  40.4 &
  \multicolumn{1}{c|}{40.9} &
  43.4 &
  44.2 \\
\multicolumn{1}{c|}{} &
  \multicolumn{1}{c|}{WCR} &
  34.5 &
  \multicolumn{1}{c|}{35.6} &
  41.5 &
  \multicolumn{1}{c|}{42.4} &
  40.6 &
  \multicolumn{1}{c|}{41.1} &
  44.3 &
  45.1 \\
\multicolumn{1}{c|}{} &
  \multicolumn{1}{c|}{HLATR} &
  \textbf{35.0} &
  \multicolumn{1}{c|}{\textbf{36.0}} &
  \textbf{42.0} &
  \multicolumn{1}{c|}{\textbf{42.9}} &
  \textbf{42.0} &
  \multicolumn{1}{c|}{\textbf{42.5}} &
  \textbf{44.5} &
  \textbf{45.2} \\ \midrule
\multicolumn{1}{c|}{\multirow{3}{*}{BERT$_{\text large}$}} &
  \multicolumn{1}{c|}{Reranking} &
  35.4 &
  \multicolumn{1}{c|}{36.3} &
  41.0 &
  \multicolumn{1}{c|}{42.0} &
  40.6 &
  \multicolumn{1}{c|}{41.1} &
  44.2 &
  45.0 \\
\multicolumn{1}{c|}{} &
  \multicolumn{1}{c|}{WCR} &
  35.6 &
  \multicolumn{1}{c|}{36.6} &
  42.0 &
  \multicolumn{1}{c|}{42.9} &
  40.8 &
  \multicolumn{1}{c|}{41.3} &
  45.2 &
  45.9 \\
\multicolumn{1}{c|}{} &
  \multicolumn{1}{c|}{HLATR} &
  \textbf{35.9} &
  \multicolumn{1}{c|}{\textbf{36.7}} &
  \textbf{42.4} &
  \multicolumn{1}{c|}{\textbf{43.4}} &
  \textbf{42.2} &
  \multicolumn{1}{c|}{\textbf{42.8}} &
  \textbf{45.6} &
  \textbf{46.3} \\ \midrule
\multicolumn{1}{c|}{\multirow{3}{*}{RoBERTa$_{\text base}$}} &
  \multicolumn{1}{c|}{Reranking} &
  34.9 &
  \multicolumn{1}{c|}{36.1} &
  40.8 &
  \multicolumn{1}{c|}{41.7} &
  40.6 &
  \multicolumn{1}{c|}{41.0} &
  44.3 &
  45.1 \\
\multicolumn{1}{c|}{} &
  \multicolumn{1}{c|}{WCR} &
  35.2 &
  \multicolumn{1}{c|}{36.3} &
  42.0 &
  \multicolumn{1}{c|}{42.9} &
  41.0 &
  \multicolumn{1}{c|}{41.6} &
  45.1 &
  45.8 \\
\multicolumn{1}{c|}{} &
  \multicolumn{1}{c|}{HLATR} &
  \textbf{35.5} &
  \multicolumn{1}{c|}{\textbf{36.5}} &
  \textbf{42.5} &
  \multicolumn{1}{c|}{\textbf{43.5}} &
  \textbf{42.0} &
  \multicolumn{1}{c|}{\textbf{42.4}} &
  \textbf{45.4} &
  \textbf{46.1} \\ \midrule
\multicolumn{1}{c|}{\multirow{3}{*}{RoBERTa$_{\text large}$}} &
  \multicolumn{1}{c|}{Reranking} &
  36.2 &
  \multicolumn{1}{c|}{37.2} &
  41.6 &
  \multicolumn{1}{c|}{42.6} &
  40.8 &
  \multicolumn{1}{c|}{41.2} &
  45.2 &
  45.9 \\
\multicolumn{1}{c|}{} &
  \multicolumn{1}{c|}{WCR} &
  36.4 &
  \multicolumn{1}{c|}{37.4} &
  42.9 &
  \multicolumn{1}{c|}{43.9} &
  41.2 &
  \multicolumn{1}{c|}{41.6} &
  45.7 &
  46.4 \\
\multicolumn{1}{c|}{} &
  \multicolumn{1}{c|}{HLATR} &
  \textbf{36.8} &
  \multicolumn{1}{c|}{\textbf{37.8}} &
  \textbf{43.7} &
  \multicolumn{1}{c|}{\textbf{44.5}} &
  \textbf{42.3} &
  \multicolumn{1}{c|}{\textbf{42.7}} &
  \textbf{45.9} &
  \textbf{46.6} \\ \bottomrule
\end{tabular}
}
\end{table*}

\begin{table}[]
\centering
\caption{Hyperameters for HLATR, Ranker for Passage Dataset (Ranker(P)) and Document (Ranker(D)), lr represents learning rate, $k$ is the number of hard negatives, $n$ is the number of transformer layers, head is the attention head for each transformer layer, $d$ is the dimension of $v$.}
\label{tab:hyper}
\begin{tabular}{c|c|c|c}
\toprule
 & HLATR & Rerenk(P) & Ranker(D)\\ \midrule
lr             & 1e-3 & 1e-5 & 1e-5\\
batch size     & 1024 & 256 & 128\\
epoch          & 40   & 3 & 3\\ 
k              & 100  & 8 & 4 \\
n              & 4    & 12/24 & 12/24\\
head           & 2    & 12 & 12\\
d              & 128  & 768/1024 & 768/1024 \\ \bottomrule
\end{tabular}
\end{table}

\noindent \textbf{Implementation Details} To verify the efficiency and robustness of the HLATR module, we implement the experiments with different retrieval and reranking models. For the retrieval stage, we report results based on both sparse and dense retrieval methods. Specifically, we choose the BM25 method as the sparse retrieval model, and two state-of-the-art dense retrieval models coCondenser~\cite{gao2021unsupervised} and ANCE~\cite{xiong2021approximate} for passage and document retrieval respectively. For the reranking stage, we leverage the commonly used reranking method based on the pre-trained language model as presented in~\cite{nogueira2019passage,gao2021rethink}. However, we conduct experiments based on different pre-trained language models (BERT$_{\text base}$, BERT$_{\text large}$, RoBERTa$_{\text base}$, RoBERTa$_{\text large}$) to verify the robustness of HLATR. 

We implement the BM25 retrieval process by the pyserini toolkit~\cite{lin2021pyserini}. For ANCE\footnote{https://github.com/microsoft/ANCE} and coCondenser\footnote{https://github.com/luyug/Condenser} methods, we use the publicly released checkpoint by authors. Following previous studies, for the passage dataset, we use the body field as the document text. For the document dataset, we combine the url, title and body field as the final document text. The maximum token length is set to $128$ and $512$ respectively. Detailed hype-parameters used in our experiments can be found in Table \ref{tab:hyper}.

\vspace{0.15cm}
\noindent \textbf{Baselines} We compare HLATR with the conventional retrieve-then-reranking models and the simple Weighted Combination of two-stage Ranking models (WCR). For each query-document pair, the WCR method directly weights combined the relevance scores produced by the retrieval stage reranking stage as the final ranking score:
\begin{equation}
\alpha f(E^Q(q), E^D(d)) + (1-\alpha)f(E^R(q,d)),
\end{equation}
where $\alpha \in (0,1)$.

\subsection{Results}
Table~\ref{tab:res} presents the final results of our method for both passage and document datasets. We highlight the best performances of each metric in bold. We find that our approach achieves the best performance among all the baselines.

Our model improves the text retrieval performances by a large margin compared to the conventional retrieve-then-reranking two-stage architecture. Specifically, the HLATR enhanced text retrieval systems improve the MRR@10 metric on passage and document datasets from $40.1$ to $42.0$ and $43.4$ to $44.5$ respectively under the dense retrieval and BERT-base reranking setting. This observation illustrates the effectiveness of combining features from the previous retrieval and reranking stage to improve the final ranking performance.

We can also observe that HLATR achieves consistent improvement whether using sparse or dense retrieval methods. We train the reranking models based on different pre-trained language models. Firstly, whether for the BERT or RoBERTa model, the performance of the large model is stably better than that of the base model, which is consistent with the results of previous studies~\cite{gao2021rethink,ma2021b}. Moreover, based on different reranking models, the improvement of the HLATR model is consistent. The above experimental result strongly depicts the robustness of our HLATR model. 

For each experimental setting, we also report the results of the WCR method. We can see that the WCR method can effectively improve the retrieval performance, which once again proves the effectiveness of the motivation in coupling retrieval and reranking features. However, we can also observe that the effect of the WCR model still lags behind the HLATR model, which further illustrates that HLATR is a model advisable architecture for large-scale text retrieval task from the perspective of coupling features.

\subsection{Discussion and Analysis}
\label{sec:discussion}

\noindent \textbf{Multi-stage Feature Coupling} We further evaluate the influence of the feature coupling mechanism in HLATR. The experimental results are presented in Table \ref{tab:ablation}. ``w/o retrieval'' denotes the setting that removes the retrieval feature and only uses the reranking stage feature as input for HLATR. Similarly, ``w/o reranking'' notes the setting that we remove the reranking stage feature for HLATR. Unlike the ``w/o retrieval'' setting, only adopting ``position embedding'' as input for the transformer encoder is extra-conventional. Therefore, we replace the reranking stage feature vector with the query embedding and document embedding generated by the dense retrieval model. We tried several methods to concatenate these two embeddings and we find that the element-wise multiply performs best. So we only report the result based on the element-wise multiply concatenation method. 

From Table \ref{tab:ablation}, we can observe that when the input feature of HLATR only contains one of the retrieval or reranking features alone, the overall ranking performance is still slightly improved. For example, the ``w/o reranking'' setting improves the MRR@10 value from $38.3$ to $39.0$ compared to the simple retrieval performance. Notably, the coupled feature input will lead to an improvement with a larger margin. This experimental result shows that coupling multi-stage features can indeed improve the final reranking performance.

\begin{table}[]
\centering
\caption{Ablation experiments result on MS MARCO Passage dataset. Dense model for retrieval stage and BERT$_{base}$ model for reranking stage.}
\label{tab:ablation}
\begin{tabular}{@{}c|cc@{}}
\toprule
\multirow{2}{*}{Model} & \multicolumn{2}{c}{Passage} \\ \cmidrule(l){2-3} 
                       & MRR@10       & MRR@100      \\ \midrule
Retrieval              & 38.3         & 39.4         \\
Reranking                 & 40.8         & 41.8         \\
HLATR                   & {\bf 42.6}         & {\bf 43.5}         \\
w/o retrieval      & 41.2         & 42.2         \\
w/o ranker         & 39.0         & 39.9         \\
 \bottomrule
\end{tabular}
\end{table}

\begin{table}[]
\centering
\caption{Experiment result on MS MARCO Passage dataset of HLATR and its two variants.}
\label{tab:exp-variants}
\begin{tabular}{@{}c|cc@{}}
\toprule
\multirow{2}{*}{Model} & \multicolumn{2}{c}{Passage} \\ \cmidrule(l){2-3} 
                       & MRR@10       & MRR@100      \\ \midrule
HLATR                   & {\bf 42.6}         & {\bf 43.5}         \\
WCR         & 41.8         & 42.8         \\
HLATR$_{linear}$            & 42.4         & 43.3         \\
HLATR$_{bce}$               & 41.7         & 42.6         \\ \bottomrule
\end{tabular}
\end{table}

\vspace{0.15cm}
\noindent \textbf{Architecture analysis of HLATR} To better analyze the advantages of our model in tackling the list aware text reranking problem, we conduct experiments with two variants:

\noindent 1) HLATR$_{linear}$: In this model, we replace the multi-layer transformer with a two-layer MLP model with ReLU as the activation function.

\noindent 2) HLATR$_{bce}$: In this model, we replace the original contrastive loss with binary cross-entropy loss, which can be formulated as:
\begin{align*}
    & \mathcal{L}_{bce}(q,D_r^{'}) = \frac{1}{Z}\sum_i^Z(y_i\log(\sigma(score(q,d_i))) \\
    &+(1-y_i)\log(1-\sigma(score(q,d_i))),
\end{align*}
where $y_i$ is the $\{0,1\}$ label indicates whether this document is relevant to the query and $\sigma$ denotes the sigmoid activation function.

The experiment result is presented in table \ref{tab:exp-variants}, from which we can see that with just a $2$-layer MLP encoder, HLATR can still outperform the WCR method significantly. However, compared with the HLATR model that uses the transformer structure as encoder, the performance of HLATR$_{linear}$ is degraded. We can infer that for such a list aware reranking task, the transformer model has a better ability in modeling the mutual associations between different documents in the input sequence.

Further, by comparing the performance of the HLATR$_{bce}$ model and the original model, we can conclude that the contrastive learning objective is more proper for the text ranking learning process. The essential reason is that the text reranking task is an optimization problem with imbalanced data. In the input sequence, the number of negative samples is much larger than the number of positive samples, and the loss of contrastive learning is more advantageous in dealing with such a data imbalance scenario.

\begin{table}[]
\centering
\caption{Time cost for different reranking models and HLATR model for per 1000 queries. The time cost for RoBERTa$_{base}$ and RoBERTa$_{large}$ is similar with the BERT$_{base/large}$ model. So we only report the result of BERT models.}
\label{tab:time}
\begin{tabular}{c|c|c}
\toprule
Model      & Passage   & Document   \\ \midrule
BERT$_{\text base}$        & 600ms         & 800ms          \\ \midrule
BERT$_{\text large}$      & 900ms          & 1200ms         \\ \midrule
HLATR        & 2ms         & 2ms          \\ \bottomrule
\end{tabular}
\end{table}

\vspace{0.15cm}
\noindent \textbf{Computational Cost of HLATR} To test the efficiency of HLATR, here we report the inference time cost for different kinds of ranking models and the HLATR model. All experiments are conducted on NVIDIA Tesla 32G V100. 

In Table~\ref{tab:time}, we can see that the time cost of HLATR is significantly smaller than the PTM based reranking models. On the passage reranking dataset, HLATR is $300$/$450$ times more computationally efficient. Therefore, although we add an extra reranking stage, it only adds a little time cost compared with the whole retrieve-then-reranking process due to its small size of parameters and effective way to calculate the relevance score.

\begin{figure}[h]
    \centering
    \resizebox{0.9\columnwidth}{!}{
    \includegraphics{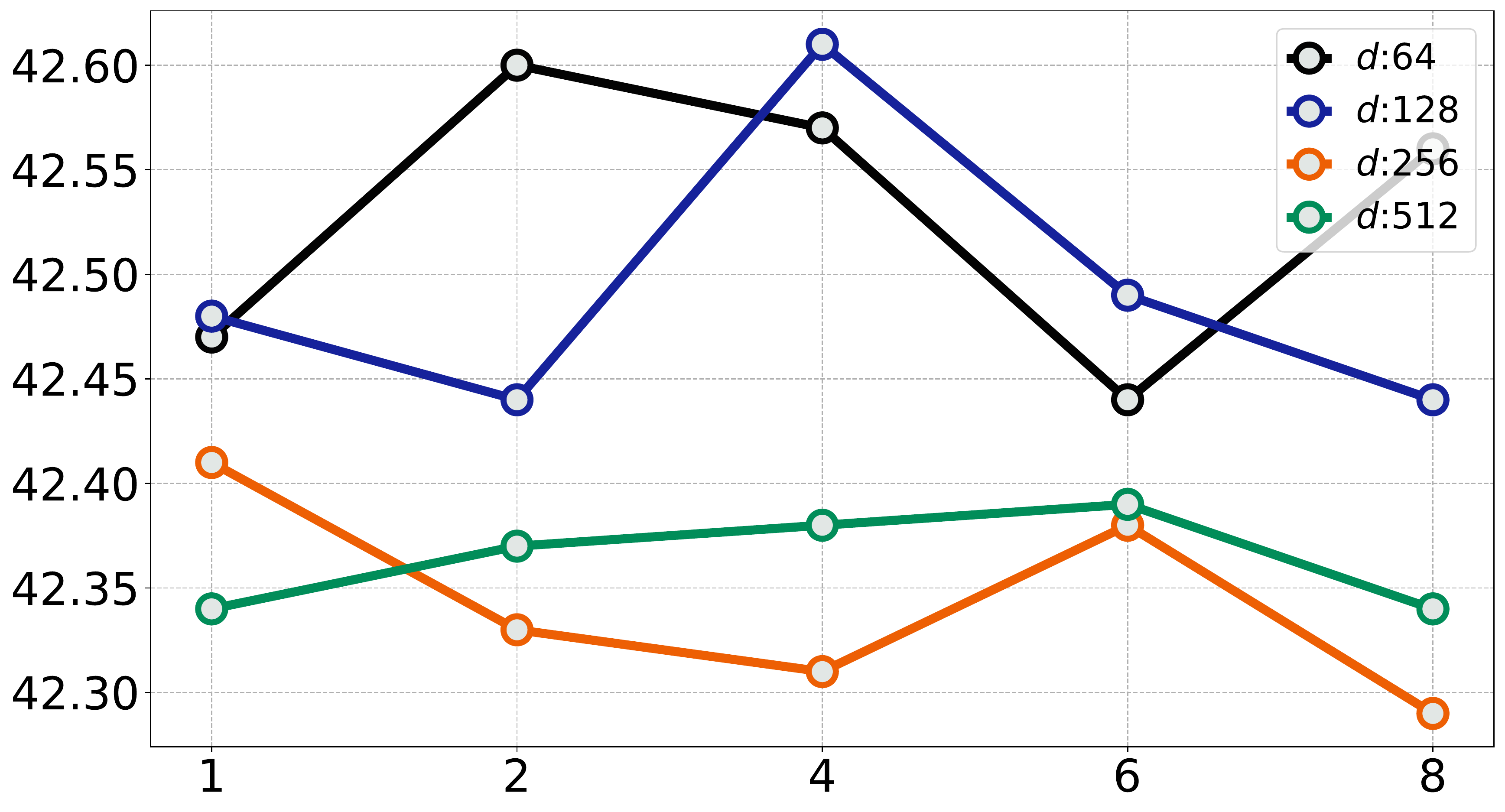}
    }
    \caption{The impact of Transformer architecture with different dimensions of hidden states $d$ and number of layers. The horizontal axis represents the number of transformer layers and the vertical axis represents the MRR@10 metric.}
    \label{fig:analysis}
\end{figure}

\noindent \textbf{Hyper-parameters of Transformer} In this part, we wish to explore how the architecture of the transformer-based encoder affects the ranking performance. In Fig~\ref{fig:analysis}, we compare the MRR@10 metric with different settings of number of transformer layers and its hidden dimension $d$ on the Passage Ranking dataset. We can see that the performance increases first and drop later when the number of layer increases. The larger $d$ is, the inflection point comes earlier. A small dimension size like $128$ leads to a better performance than a large dimension size like $512$. This can be attributed that the aim of HLATR is only designed to make use of the first two-stages features produced by the large-scale pre-trained model. So that it does not need a big number of parameters. Consequently, we need to balance the ``width'' and the ``depth'' of HLATR to achieve the best performance.

\section{Conclusion}
In this paper, we intuitively propose a unified hybrid list aware transformer reranking (HLATR) model via coupling retrieval and reranking features for multi-stage text retrieval task. The HLATR model is lightweight and can be easily applied to existing systems. Experiments on two large scale text retrieval datasets show our framework significantly outperforms previous baselines. 

\bibliography{hlatr}

\end{document}